\documentclass[conference]{IEEEtran}
\IEEEoverridecommandlockouts
\usepackage{cite}
\usepackage{amsmath,amssymb,amsfonts}
\usepackage{graphicx}
\usepackage{textcomp}
\usepackage{xcolor}
\def\BibTeX{{\rm B\kern-.05em{\sc i\kern-.025em b}\kern-.08em
    T\kern-.1667em\lower.7ex\hbox{E}\kern-.125emX}}

\usepackage{todonotes}
\usepackage{enumitem}
\usepackage{amsfonts}
\usepackage{amsmath}
\usepackage{algorithm}
\usepackage{algpseudocode}% http://ctan.org/pkg/algorithmicx
\usepackage{hyperref}
\usepackage{multirow}
\usepackage{multicol}
\usepackage{subcaption}
\usepackage[capitalize]{cleveref}
\usepackage{epsfig}
\usepackage{amssymb}
\usepackage{graphicx}
\usepackage{placeins}
\usepackage{tabularx}
\usepackage{booktabs}
\usepackage{epstopdf}

\makeatletter
\renewcommand{\ALG@beginalgorithmic}{\small\setlength{\ALG@tlm}{-0.5em}}
\makeatother

\newcommand{\ba}[1]{\begin{array}{#1}}

\newcommand{\ea}{\end{array}}

\newcommand{\beq}[1]{\begin{equation}\label{#1}}
\newcommand{\eeq}{\end{equation}}

\usepackage{color}

\begin{document}

\title{Adaptive Sketching Based Construction of H2 Matrices on GPUs}

%
% The "author" command and its associated commands are used to define
% the authors and their affiliations.
% Of note is the shared affiliation of the first two authors, and the
% "authornote" and "authornotemark" commands
% used to denote shared contribution to the research.
\author{
	\IEEEauthorblockN{
		Wajih Halim Boukaram\IEEEauthorrefmark{1}, 
		Yang Liu\IEEEauthorrefmark{2}, 
		Pieter Ghysels\IEEEauthorrefmark{3}, 
		Xiaoye Sherry Li\IEEEauthorrefmark{4}
	}
	\IEEEauthorblockA{
		\textit{Lawrence Berkeley National Laboratory}, \textit{1 Cyclotron Road, Berkeley, CA, USA}\\
	}
	\IEEEauthorblockA{\IEEEauthorrefmark{1}wajih.boukaram@lbl.gov, 
		\IEEEauthorrefmark{2}liuyangzhuan@lbl.gov, 
		\IEEEauthorrefmark{3}pghysels@lbl.gov, 
		\IEEEauthorrefmark{4}xsli@lbl.gov}
}

\maketitle

\begin{abstract}
We develop a novel linear-complexity bottom-up sketching-based algorithm for constructing a $\mathcal{H}^2$ matrix, and present its high performance GPU implementation. 
The construction algorithm requires both a black-box sketching operator and an entry evaluation function.
%In addition to the CPU implementation, 
%We present a GPU algorithm and implementation of .  
The novelty of our GPU approach centers around the design and implementation of the above two operations in batched mode on GPU with accommodation for variable-size data structures in a batch.
The batch algorithms minimize the number of kernel launches and maximize the GPU 
throughput. 
% both of which are accelerated by batched GPU implementations. 
When applied to covariance matrices, volume IE matrices and $\mathcal{H}^2$ update operations, our proposed GPU implementation achieves up to $13\times$ speedup over our CPU implementation, and up to $1000\times$ speedup over an existing GPU implementation of the top-down sketching-based algorithm from the H2Opus library. It also achieves a $660\times$ speedup over an existing sketching-based $\mathcal{H}$ construction algorithm from the ButterflyPACK library. Our work represents the first GPU implementation of the class of bottom-up sketching-based $\mathcal{H}^2$ construction algorithms. 

%This paper presents a bottom-up partially black-box sketching-based construction algorithm for $\mathcal{H}^2$ matrices on GPUs.
%The algorithm is well-suited for compressing operators in integral equations and frontal matrices in sparse direct solvers, and is the   first GPU implementation of the class of bottom-up sketching-based $\mathcal{H}^2$ construction algorithms. 
%Compared with existing top-down sketching-based construction algorithms on GPUs, the proposed algorithm requires much fewer samples and is asymptotically superior.
\end{abstract}



%\xslnote{Abstractt due: 3/26; Full paper due: 4/2}

\section{Introduction}

% \xslnote{
% \begin{itemize}
%     \item Why do $\mathcal{H}^2-matrix$? \\
%     (strong admissibility; nested basis for linear time)
%     \item deficiencies of previous construction algorithms (challenges)\\
%     (this paper only deals with construction, no factorization)
%     \item why randomized sketching (with adaptivity)?
    
%     Sell this algorithm  mainly useful in sparse factorization, e.g., efficient extend-add to form Schur complement, simpler data structure
%     \item Summary of our contributions\\
%     GPU
% \end{itemize}
% }

% %\section{Related Work}

% \xslnote{
%   \begin{itemize}
%     \item H2OPUS~\cite{zampini2022h2opus}
%     \item Ying's paper \cite{minden2017recursive}
%     \item randomized strong recursive skeletonization~\cite{Yesypenko} (Yang/Sherry on it)
%       \item Miaomiao Ma's paper~\cite{Ma2019} (Yang/Sherry on it)\\
%       factoization only; serial
%     \item H2Lib\footnote{\url{http://www.h2lib.org/}}\\
%     shared-memory?
%     \item H2Pack~\cite{H2pack}\\
%     kernel-dependent? proxy-points; shared-memory
%     \item distributed $\mathcal{H}^2$~\cite{Borm2023,liang2024on,ma2022scalable}\\
    
% \end{itemize}
% }

Many large-scale dense matrices from scientific and engineering applications exhibit low-rank structure after proper hierarchical matrix partitioning. 
Such low-rank structure can be exploited by hierarchical matrix techniques to enable fast matrix-vector multiplication and matrix inversion in quasi-linear time. Examples include integral equation methods for acoustics, electromagnetics~\cite{hackbusch1999sparse,bebendorf2003existence,borm2003introduction}, Stokes flows \cite{zhang2022fast} and charged particle systems \cite{tu2024hierarchical}, differential equation-based PDE solvers ~\cite{xia2013randomized,ghysels2017robust}, machine learning methods like kernel ridge regression \cite{chavez2020scalable} and Gaussian processes \cite{ambikasaran2015fast}, and various other structured matrices, e.g., Toeplitz and Cauchy~\cite{chandrasekaran2008superfast}. 

There exists a broad family of hierarchical matrix techniques, including the $\mathcal{H}$/$\mathcal{H}^2$ formats~\cite{hackbusch1999sparse,bebendorf2003existence,borm2003introduction}, the hierarchically off-diagonal low-rank format (HODLR)~\cite{ambikasaran2013mathcal}, the hierarchically semi-separable format (HSS)~\cite{chandrasekaran2007fast} or hierarchically block separable format (HBS)~\cite{gillman2012direct}, the inverse fast multipole method (IFMM) \cite{coulier2017inverse} and the hierarchical interpolative factorization (HIF) algorithms \cite{l2016hierarchical}. 
These formats can be characterized by the so-called admissibility condition which determines how much separated interaction can be low-rank compressed. The optimal choice of hierarchical format depends on the particular application, including the dimensionality and discretization scheme. 
For high-dimensional problems, weak-admissibility-based formats such as HODLR, HSS, HBS typically cannot achieve quasi-linear complexities with the exception of HIF, which however only has been demonstrated with regular-grid-based discretization. 
On the other hand, strong-admissibility-based formats can attain quasi-linear (e.g. $\mathcal{H}$) and linear complexities (e.g. $\mathcal{H}^2$ and IFMM) for high-dimensional problems. That being said, they usually show larger prefactors and/or pose challenges for scalable parallel implementations.
     
This paper focuses on efficient algorithms for the construction of the $\mathcal{H}^2$ format. 
Just like the other hierarchical matrix formats, a $\mathcal{H}^2$ matrix can be efficiently constructed by assuming that (a) any matrix entry can be computed quickly on-the-fly \cite{borm2003introduction,H2pack,zampini2022h2opus,Borm2023} 
or (b) a fast black-box sketching operator is available. Here (a) is commonly encountered for compressing forward operators in integral equations and kernel matrices. Existing codes include HLIBpro \cite{borm2003introduction,Borm2023}, H2Pack \cite{H2pack}, ASKIT \cite{askit}, GOFMM \cite{gofmm}, and GPU implementations like H2Opus \cite{zampini2022h2opus} and hmglib \cite{hmatrixgpu}.
They typically leverage adaptive cross approximation, proxy surface, or preselected skeletons to construct the $\mathcal{H}^2$ matrix. On the other hand, (b) is often encountered for compressing frontal matrices in sparse multifrontal solvers, trace estimation in Bayesian optimization, or low-rank updating an existing $\mathcal{H}^2$ matrix. 
Unlike (a), there exist fewer known sketching-based algorithms and implementations for (b), which include the top-down peeling algorithms \cite{lin2011fast,levitt2022randomized,zampini2022h2opus} and the more recent bottom-up algorithm \cite{Yesypenko}. 

We focus on algorithms based on assumption (b) in this paper. 
We propose a partially black-box sketching-based $\mathcal{H}^2$ construction algorithm requiring fewer samples compared to existing algorithms and describe an efficient GPU implementation, particularly useful for accelerating $\mathcal{H}^2$ arithmetic in sparse multifrontal solvers or Schur complement-based updates. 
It is worth mentioning that once the $\mathcal{H}^2$ matrix has been constructed, efficient (i.e., low-prefactor) inversion algorithms have also been recently developed \cite{minden2017recursive,Ma2019} and parallelized \cite{liang2024on,ma2022scalable}.
But there is no GPU algorithm for inversion.
This current paper describes the construction phase for $\mathcal{H}^2$ on GPU. In a future paper, we will describe the GPU algorithm for $\mathcal{H}^2$ inversion.  

Our main contributions can be summarized as follows:
\begin{itemize}
\item We develop a novel partially black-box $\mathcal{H}^2$ matrix construction algorithm with linear complexity, which extends the bottom-up algorithm in \cite{ghysels2017robust,martinsson2011fast} from weakly-admissible HSS to strongly-admissible $\mathcal{H}^2$ and permits adaptive sketching. 
Compared with the top-down algorithms in \cite{levitt2022randomized,zampini2022h2opus}, the proposed algorithm requires much fewer samples and is asymptotically faster. 
\item We develop a GPU implementation of the construction algorithm relying on batched dense linear algebra kernels and batched entry extraction routines. This represents the first parallel GPU implementation of the partially black-box or fully black-box \cite{Yesypenko} bottom-up construction algorithms. 
\item We demonstrate the computational efficiency of the proposed algorithm by compressing integral equations and covariance matrices, and recompressing $\mathcal{H}^2$ matrices with low-rank updates. Our GPU implementation achieves up to $13\times$ speedup over our CPU implementation, and up to $1000\times$ speedup over the GPU implementation of a top-down sketching-based algorithm in H2Opus. It's also worth mentioning that our CPU and GPU implementations share the same code base due to the use of Thrust. 
\end{itemize}

The remainder of the paper is structured as follows: Section~\ref{sec:preliminaries} introduces $\mathcal{H}^2$ matrices and associated preliminaries including cluster tree and interpolative decomposition. Sections~\ref{sec:sketching_construction} and~\ref{sec:gpu_implementation} describe the main contributions, the adaptive sketching construction and the high-performance GPU implementation, respectively. Performance results are reported in Section~\ref{sec:application_and_performance} and we conclude in Section~\ref{sec:conclusion}.

\section{Preliminaries}
\label{sec:preliminaries}

\subsection{Hierarchical Matrices}

\begin{figure}[t]
  \centering
  \includegraphics[width=0.6\linewidth]{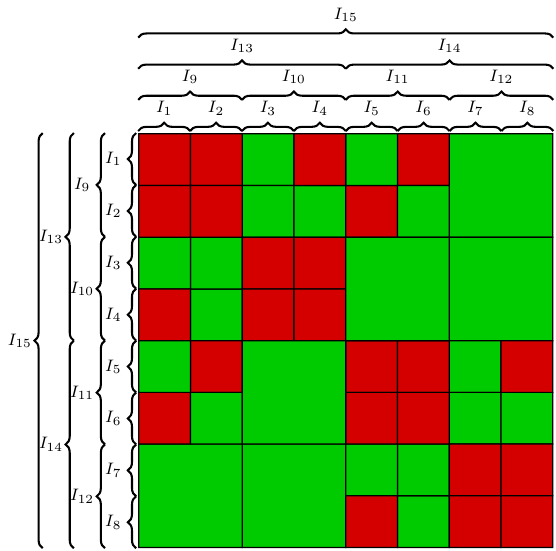}  
  \caption{The leaves of the hierarchical matrix tree forming a block partitioning of the matrix. Red blocks represent inadmissible leaves and green blocks represent admissible blocks. Row and columns indices are hierarchically clustered into a cluster tree $I$ such that pairs of clusters define blocks within the matrix. }
  \label{fig:original_hmatrix}
\end{figure}

\begin{figure}[tbh!p]
  \centering
  \includegraphics[width=0.7\linewidth]{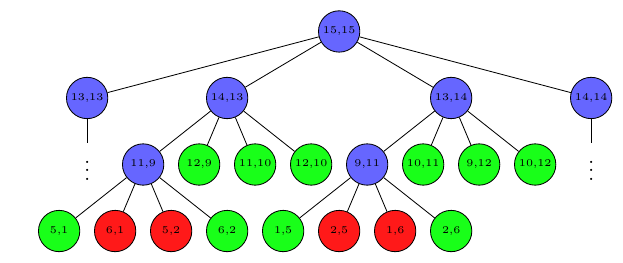}  
  \caption{The matrix tree for the hierarchical matrix in Fig. \ref{fig:original_hmatrix} representing the inadmissible blocks in blue, the admissible leaves in green and the inadmissible leaves in red. In general, the matrix tree is not a complete tree. The ellipsis represent a complete subtree of the tree and are omitted for brevity.}
  \label{fig:matrix_tree}
\end{figure}
\begin{figure}
	\centering
	\includegraphics[width=0.7\linewidth]{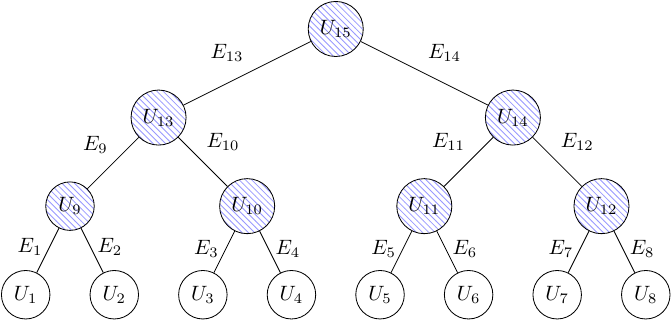}  
	\caption{The basis tree for the hierarchical matrix in Fig. \ref{fig:original_hmatrix} where leaves $U_\tau$ are stored explicitly and the shaded inner nodes are implicitly represented by the nested basis property using the transfer matrices $E$.}
	\label{fig:basis_tree}
\end{figure}

Hierarchical matrices aim to provide an efficient representation of dense matrices that are data sparse, where certain blocks within these dense matrices can be well approximated by a low-rank form. 
Many different variants of hierarchical matrices have been developed over the years, primarily differentiating themselves on the block partitioning used and the representation of the basis vectors used to approximate the blocks. 
% For a more detailed introduction to the subject, we refer the reader to [insert a good reference here]. 
The block partitioning is typically determined by first hierarchically clustering the indices of the matrix $K$ into a cluster tree $I$, and then performing a dual tree traversal on $I$. 
The traversal generates pairs of clusters $(s, t)$ that are tested against an admissibility condition that determines whether the matrix block defined by the cluster pair can be approximated well by a low-rank matrix. We consider the so-called general admissibility condition \texttt{adm} which determines the compressibility of a block based on the distance \texttt{Dist} between the bounding boxes of the cluster pair $(s,t)$ and the average of their diameters $D$:
\begin{equation}
\texttt{adm}(s,t)=1,~if~\frac{D(s) + D(t)}{2} \leq \eta \texttt{Dist}(s, t)
\end{equation}
Typically $\eta\geq1$ indicates the so-called weak admissibility and $\eta\leq0.5$ indicates the so-called strong admissibility. 
The general admissibility condition is used to perform a dual tree traversal. The traversal produces a matrix tree where each node is a cluster pair $(s, t)$ at the same level. If a cluster pair is deemed inadmissible, the dual tree traversal continues on their four children until the block defined by the pair is sufficiently small and thus stored in its original dense form. The full set of leaves of this matrix tree then define the block partitioning of the matrix. Fig.~\ref{fig:original_hmatrix} illustrates the cluster tree $I$ whose dual traversal with a general admissibility condition 
%\xslnote{define it}
produced a block partitioning of the matrix where admissible leaves are shown in green and inadmissible leaves are shown in red. 
The corresponding matrix tree is shown in Fig. \ref{fig:matrix_tree} with the 
complete subtrees $K_{13,13}$ and $K_{14,14}$ on the diagonal omitted for brevity. 
All the leaves within a level of the matrix tree can be viewed as a block sparse matrix, and one important property of hierarchical matrices is that the number of blocks in a row of each level's block sparse matrix is bounded by a constant that does not grow with the problem size. This constant is called the sparsity constant $C_{sp}$. 

More specifically, Fig.~\ref{fig:h2algo}(a)-(b) shows the block partitioning of a matrix associated with a set of $N=2^{15}$ 3D geometry points, using the admissibility parameter $\eta=0.5, 0.7$. Note that smaller $\eta$ leads to more refined partitioning of the off-diagonal blocks, and hence larger sparsity constants $C_{sp}$.

%
%\begin{algorithm}[t]
%\caption{Dual Tree Traversal}
%\raggedright
%\textbf{Input:} Cluster tree nodes $s$ and $t$, general admissibility condition \texttt{adm} and a leaf size $m$\\
%\textbf{Output:} Matrix tree and the block partitioning of the hierarchical matrix
%\label{alg:dual_traversal}
%\begin{algorithmic}[1]
%\Procedure{DualTreeTraversal}{$s$, $t$}
%	\If {\texttt{adm}($s$, $t$)}             
%		\State Matrix tree leaf $(s, t)$ is low rank
%	\ElsIf {size($s$) $\leq m$ or size($t$) $\leq m$}
%	  	\State Matrix tree leaf $(s, t)$ is dense
%	\Else
%		\For {$s_i \in $ children($s$)}
%			\For {$t_j \in $ children($t$)}
%				\State DualTreeTraversal($s_i, t_j$)
%			\EndFor
%		\EndFor
%	\EndIf
%\EndProcedure
%\end{algorithmic}
%\end{algorithm}
%

Let us denote the block of a matrix $K$ determined by a cluster pair $(s, t)$ as $K_{s,t}$, and the set of clusters that form inadmissible pairs with a cluster $\tau$ as $\mathcal{N}_\tau$. The set of clusters (1) that form admissible pairs with cluster $\tau$ and (a) whose parents form inadmissible pairs with parent of $\tau$, are denoted as $\mathcal{F}_\tau$. These notations are summarized in Table \ref{tbl:notation}. We also use MATLAB notation when convenient.
Now that we've determined the desired block partitioning, we would like to represent the admissible blocks $K_{s,t}$ in a low-rank form. The $\mathcal{H}$-matrix variant represents each $m \times n$ block $K_{s,t}$ of rank $k$ as the outer product $K_{s,t} = U_{s,t} V_{s,t}^T$, where $U_{s,t}$ and $V_{s,t}$ are $m \times k$ and $n \times k$ matrices respectively, leading to $O(n \log n)$ storage complexity for the matrix. 
On the other hand, the $\mathcal{H}^2$-matrix variant use a nested basis to achieve $O(n)$ storage complexity.
Instead of storing independent $U$ and $V$ matrices for each block, $\mathcal{H}^2$-matrices use a common basis for the block rows/columns defined by each cluster in the cluster tree, introducing a smaller $k \times k$ coupling matrix $B$ for each block and representing each block as $K_{s,t} = U_{s} B_{s,t} V_{t}^T$. For simplicity, we assume $K$ is symmetric and real-valued in the rest of this paper, leading to $V_{t}=U_{s}^T$. However, our algorithm can be easily extended to un-symmetric or complex-valued matrices. 

The basis for leaf nodes in the cluster tree are stored explicitly, and the basis for an inner node $\tau$ of the cluster tree is defined in terms of the basis of its children $\tau_1$ and $\tau_2$ using transfer matrices $E$, resulting in a nested basis:

\begin{equation}
    \label{eqn:nested_basis}
    U_\tau = \begin{bmatrix}
    U_{\tau_1} & \\
     & U_{\tau_2}
    \end{bmatrix} \begin{bmatrix}
    E_{\tau_1} \\
    E_{\tau_2}
    \end{bmatrix}
\end{equation}

Fig. \ref{fig:basis_tree} shows a basis tree for the clusters in Fig. \ref{fig:original_hmatrix}, where the clear nodes at the leaf level are stored explicitly and the shaded internal nodes are represented implicitly using the nested basis property. 

\subsection{Interpolative Decomposition}
\label{sec:interpolative_decomposition}
The interpolative decomposition (ID) aims to compute a factorization of an $m \times n$ matrix $A$ such that $A$ can be approximated as a linear combination of a set $S$ of selected columns: $A \approx A(:, S) X$, where the rank $k=\textrm{card}(S)$ is usually selected to satisfy some approximation threshold $\epsilon$. We refer to this ID as the column ID. The column ID can be computed using the column pivoted QR decomposition, where a column permutation $P$ of $A$ is factored into an orthogonal factor $Q$ and a triangular factor $R$: $AP = QR$. The column ID can then be computed as follows:
\begin{align} 
    AP & = QR = \begin{bmatrix} Q_1 & Q_2 \end{bmatrix} \begin{bmatrix}
        R_1 & R_2 \\\nonumber
        0 & R_3
    \end{bmatrix} \\\nonumber 
    & = Q_1 \begin{bmatrix} R_1 & R_2 \end{bmatrix} + Q_2 \begin{bmatrix} 0 & R_3 \end{bmatrix} \\
    & \approx Q_1 R_1 \begin{bmatrix} I & R_1^{-1}R_2 \end{bmatrix} = A(:, S) \begin{bmatrix} I & T \end{bmatrix}
\end{align}
By discarding the lower right triangular factor $R_3$ when its norm becomes small enough to guarantee the approximation threshold $\epsilon$ for $A$, we can obtain the interpolation matrix $T = R_1^{-1}R_2$. Similarly, the row ID for A is defined as $PA=\begin{bmatrix}I&T\end{bmatrix}^TA(S,:)$, which is typically computed via the column ID of $A^T$. When referring to matrix column or row indices, we will refer to the set of selected indices $S$ as the skeletonization indices and the remaining unselected indices as the redundant indices $R$\footnote{$R$ is used from this point onward to denote the redundant indices.}.

\begin{table}
\caption{Notation}
\label{tbl:notation}
\begin{tabularx}{\linewidth}{p{0.05\textwidth}X}
\toprule
  $\tau$    & A node in the cluster tree \\
  $U_\tau$  & Basis matrix for a leaf node \\
  $E_\tau$  & Transfer matrix for an inner node  \\
  $B_{s,t}$ & Coupling matrix for admissible node pair $(s,t)$  \\
  $D_{s,t}$ & Dense leaf matrix for inadmissible node pair $(s,t)$ \\
  $\mathcal{N}_\tau$  & Set of clusters that form inadmissible pairs with $\tau$  \\
  $\mathcal{F}_\tau$  & Set of clusters that form admissible
pairs with $\tau$ \\
\bottomrule
\end{tabularx}
\end{table}

\begin{algorithm}[th!]
	\caption{Proposed $\mathcal{H}^2$ construction algorithm for a matrix $K$ based on sketching (permutation matrices are not shown)}
	\label{alg:H2-sketch}
	\raggedright
	\textbf{Input:} Sample block size $d$, a hierarchical partitioning of the blocks of the matrix of $L$ levels, a relative compression tolerance $\epsilon$, a black-box function $Y=K_{blk}(\Omega)$ that can compute $Y=K\Omega$ with a random matrix $\Omega\in \mathbb{R}^{N\times d}$ in $O(Nd)$ time, and a function to evaluate any subblock $K_{s,t}$. \\
	\textbf{Output:} Skeletonization indices $\tilde{I}_\tau$ for each node $\tau$. $\mathcal{H}^2$ matrix $K_\mathcal{H}$ with, for each node $\tau$, $U_\tau$, $D_{\tau,b}$ ($b\in\mathcal{N}_\tau$), $B_{\tau,b}$ ($b\in\mathcal{F}_\tau$) at the leaf level, and $E_{\tau_1}$,$E_{\tau_2}$, $B_{\tau,b}$ ($b\in\mathcal{F}_\tau$) at higher levels.
	\begin{algorithmic}[1]
		\State $Y=K_{blk}(\Omega)$ with a random $\Omega \in \mathbb{R}^{N\times d}$\Comment{\textcolor{teal}{batchedRand}}\label{line:rand} 
		\For{level $l=1,\ldots,L$} 
		\If{$l = 1$}\Comment{Leaf node}		
		\For{node $\tau$ at level $l$} 
		
		\State ${\Omega}^1_\tau=\Omega(I_\tau, :),{Y}^1_\tau=Y(I_\tau, :)$
		\EndFor	  		 
		\For{node $\tau$ at level $l$} 
		
		\State$D_{\tau,b}=K({I}_\tau,{I}_b)$ $\forall b\in\mathcal{N}_\tau$, \Comment{\textcolor{teal}{batchedGen}}\label{line:D}
		\State$Y_\tau^{\mathrm{loc}}= {Y}^1_\tau-\sum_{b\in\mathcal{N}_\tau}D_{\tau,b}{\Omega}^1_b$\label{line:yloc_leaf} \Comment{\textcolor{teal}{batchedBSRGemm}}
		\EndFor	  	
		\textcolor{gray}{		      \While {$\exists$ $\tau$ non-converged (via QR of $Y_\tau^{\mathrm{loc}}$)} \label{line:convergence_leaf}  
			\State \hspace{-0.35cm}$\bar{Y}=K_{blk}(\bar{\Omega})$ with a random $\bar{\Omega} \in \mathbb{R}^{N\times d}$\hspace{-0.1cm}\Comment{\textcolor{teal}{batchedRand}}\label{line:rand_bar_leaf} 
			\State \hspace{-0.35cm}$Y_\tau^{\mathrm{loc}}, \Omega_\tau^l, \forall \tau$ at $l$ = updateSamples($\bar{Y}, \bar{\Omega}, l$)\label{line:sampleupdate_leaf}
			\EndWhile   }		    
		
		\For{node $\tau$ at level $l$} 
		\State $Y_\tau^\mathrm{loc}=U_\tau Y_\tau^{\mathrm{loc}}(J_\tau,:)$ \label{line:ID_leaf} \Comment{ID with $\epsilon_l$. \textcolor{teal}{batchedID}}
		\State $Y^{l+1}_\tau=Y_\tau^{\mathrm{loc}}(J_\tau,:)$\label{line:update_Y_leaf}\Comment{\textcolor{teal}{batchedShrink}}
		\State $\Omega^{l+1}_\tau=U_\tau^T{\Omega}^l_\tau$\label{line:update_Omega_leaf}\Comment{\textcolor{teal}{batchedGemm}}
		\State $\tilde{I}_\tau={I}_\tau(J_\tau)$ \label{line:skeleton_leaf}\Comment{Pick the $J_{\tau}$ indices from ${I}_{\tau}$}      
		\EndFor	  		
		\Else  \Comment{Inner node}
		\For{node $\tau$ at level $l$} 
		\State Let $\nu_1$ and $\nu_2$ be the children of $\tau$
		
		\State $\bar{I}_\tau=[\tilde{I}_{\nu_1},\tilde{I}_{\nu_2}], {\Omega}^l_\tau=\begin{bmatrix}
		\Omega^l_{\nu_1}\\
		\Omega^l_{\nu_2}
		\end{bmatrix},{Y}^l_\tau=\begin{bmatrix}
		Y^l_{\nu_1}\\
		Y^l_{\nu_2}
		\end{bmatrix}$\label{line:merge}		
		\EndFor 		
		\For{node $\tau$ at level $l$} 
		
		\State \hspace{-0.2cm}$Y_\tau^{\mathrm{loc}}=Y_\tau^l-\begin{bmatrix}
		\sum_{b\in\mathcal{F}_{\nu_1}}B_{\nu_1,b}\Omega^l_{b} \\
		\sum_{b\in\mathcal{F}_{\nu_2}}B_{\nu_2,b}\Omega^l_{b}
		\end{bmatrix}$\label{line:yloc_high}\hspace{-1cm}\Comment{\textcolor{teal}{batchedBSRGemm}}
		\EndFor
		\textcolor{gray}{		      \While {$\exists$ $\tau$ non-converged (via QR of $Y_\tau^{\mathrm{loc}}$)} \label{line:convergence_high}  
			\State \hspace{-0.35cm}$\bar{Y}=K_{blk}(\bar{\Omega})$ with a random $\bar{\Omega} \in \mathbb{R}^{N\times d}$\hspace{-0.1cm}\Comment{\textcolor{teal}{batchedRand}}\label{line:rand_bar_high} 
			\State \hspace{-0.35cm}$Y_\tau^{\mathrm{loc}}, \Omega_\tau^l, \forall \tau$ at $l$ = updateSamples($\bar{Y}, \bar{\Omega}, l$)\label{line:sampleupdate_high}
			\EndWhile   }			        		
		\For{node $\tau$ at level $l$} 
		\State $Y_\tau^\mathrm{loc}=\begin{bmatrix}E_{\tau_1} \\ E_{\tau_2}\end{bmatrix} Y_\tau^{\mathrm{loc}}(J_\tau,:)$ \hspace{-1cm}\Comment{ID with $\epsilon_l$. \textcolor{teal}{batchedID}}\label{line:ID_high}
		\State $Y^{l+1}_\tau=Y_\tau^{\mathrm{loc}}(J_\tau,:)$\label{line:update_Y_high}\Comment{\textcolor{teal}{batchedShrink}}
		\State $\Omega^{l+1}_\tau=\begin{bmatrix}E_{\tau_1}^T & E_{\tau_2}^T\end{bmatrix}{\Omega}^l_\tau$\label{line:update_Omega_high}\Comment{\textcolor{teal}{batchedGemm}}
		\State $\tilde{I}_\tau=\bar{I}_\tau(J_\tau)$ \Comment{Pick the $J_{\tau}$ indices from $\bar{I}_{\tau}$}        		
		\EndFor				        		
		\EndIf
		
		\For{node $\tau$ at level $l$} 
		\State$B_{\tau,b}=K(\tilde{I}_\tau,\tilde{I}_b)$ $\forall b\in\mathcal{F}_\tau$ \Comment{\textcolor{teal}{batchedGen}}\label{line:B}
		\EndFor		    		
		
		\EndFor
	\end{algorithmic}
\end{algorithm}

\section{Sketching Construction}
\label{sec:sketching_construction}
In this section, we discuss the details of the construction of hierarchical matrices using the sketching algorithm, starting with the fixed sample version in Section \ref{sec:fixed_sample_construction} that assumes that the number of samples needed for construction is known beforehand. The proposed algorithm represents the extension of a sketching-based construction algorithm for the HSS matrix \cite{martinsson2011fast} to strongly-admissible $\mathcal{H}^2$ matrices. This algorithm is then generalized to the adaptive version in Section \ref{sec:adaptive_sample_construction} that adds additional samples as needed to satisfy a compression threshold $\epsilon$. We will assume that the matrix is symmetric to simplify the discussion, as non-symmetric matrices are a straightforward modification to the algorithm. We also assume that a hierarchical block partitioning of the matrix that would allow for low-rank compression is already computed.

%\subsection{Fixed Sampling Construction}
\subsection{Construction with Fixed Sample Size}
\label{sec:fixed_sample_construction}

Let us first assume that we know the representative rank $r$ of the hierarchical matrix that we're constructing and specify an oversampling parameter $p$ designed to assure with high probability that the total number of samples $d=r+p$ is sufficient to guarantee the construction to the required accuracy. To simplify the discussion, let us also assume that the indices of the matrix are already sorted so that all redundant indices $R$ within a cluster come before all the skeletonization indices $S$. In other words, we drop the permutation matrix $P$ in column and row IDs in what follows. The randomized construction algorithm (See Algorithm \ref{alg:H2-sketch}) requires two inputs:  (a) a black box function that computes $Y=K \Omega$ with a random set of vectors $\Omega \in \mathbb{R}^{N\times d}$, and (b) a function used to evaluate a small number of matrix entries. The algorithm uses $Y$ to recursively sketch the admissible blocks at each level of the $\mathcal{H}^2$ matrix, after subtracting out contribution to $Y$ of the inadmissible blocks at each level via direct matrix entry evaluations. Note that for now one can ignore the while loops in grey at Lines \ref{line:convergence_leaf} and \ref{line:convergence_high}, which will be explained in Section \ref{sec:adaptive_sample_construction} in the context of adaptive sampling.

\subsubsection{Construction at the leaf level}

\begin{figure*}[thp!]
	\centering
	\includegraphics[width=\textwidth]{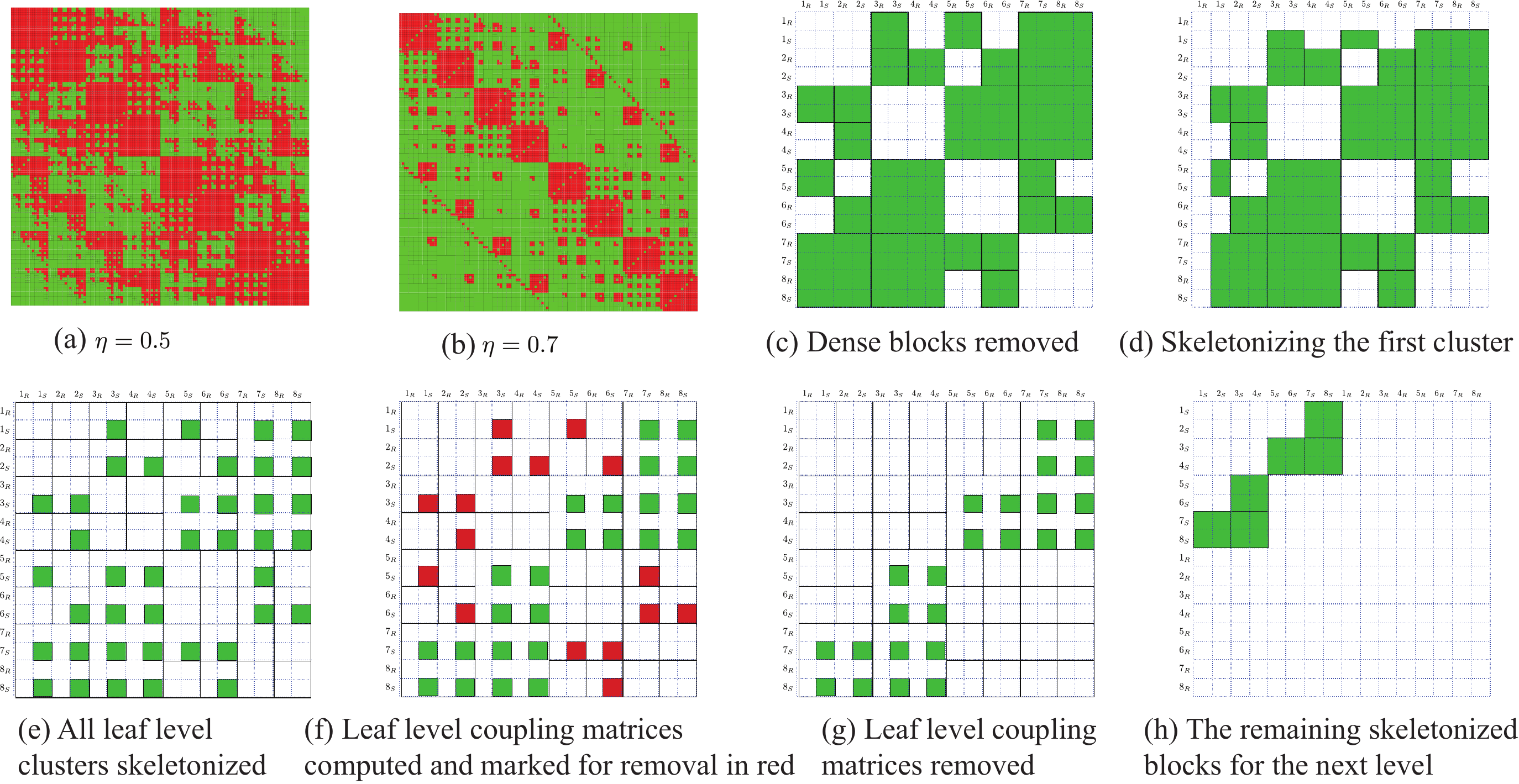}
	\caption{(a)-(b) Block partitioning of a hierarchical matrix for a 3D problem of size $N=2^{15}$ with different $\eta$. (c)-(h) The skeletonization process for the leaf level of the $\mathcal{H}^2$-matrix in Fig. \ref{fig:original_hmatrix}. }
	\label{fig:h2algo}
\end{figure*}

At the leaf level, tthe influence of the inadmissible leaf blocks is first subtracted out from the samples so that we are left with just the samples of the admissible part of the matrix. For each node $\tau$, the dense blocks $D_{\tau, b}$, $b\in\mathcal{N}_\tau$ are evaluated using the index sets of its defining clusters $I_\tau, I_b$, multiplied by the submatrices of the input vector corresponding to the cluster $t$, $\Omega^1_t = \Omega(I_t, :)$ and then subtracted from the samples submatrix $Y^1_\tau = Y(I_\tau, :)$. This is reflected by Line \ref{line:yloc_leaf} of Algorithm \ref{alg:H2-sketch}: $Y_\tau^{\mathrm{loc}}= {Y}^1_\tau-\sum_{b\in\mathcal{N}_\tau}D_{\tau,b}{\Omega}^1_b$. Fig. \ref{fig:h2algo}(c) shows the admissible blocks which contribute to each $Y_\tau^{\mathrm{loc}}$. 

Performing the row ID on the samples $Y_\tau^{\mathrm{loc}}$ for each cluster $\tau$, $Y_\tau^{\mathrm{loc}}= \begin{bmatrix} T_\tau  & I  \end{bmatrix}^TY_\tau^{\mathrm{loc}}(J_\tau,:)$, gives us an interpolation matrix $T_\tau$ that can be used as the interpolation matrix for the admissible block row/column for the cluster. The basis for cluster $\tau$ can thus be computed using the interpolation matrix as $U_\tau = \begin{bmatrix}
T_\tau & I 
\end{bmatrix}^T$. See Line \ref{line:ID_leaf} of Algorithm \ref{alg:H2-sketch}. Let us define two block unit-triangular matrices:
\begin{equation}
    W_\tau=\begin{bmatrix}
    I & -T^T_\tau\\ 0 & I
\end{bmatrix} \quad\mathrm{and}\quad  
Z_\tau=W_\tau^T=\begin{bmatrix}
    I & 0 \\ -T_\tau & I
\end{bmatrix}
\end{equation}
After scaling the block row $K(I_s,:)$ by $W_s$ and block column $K(:,I_t)$ by $Z_t$, each admissible matrix block $K(I_s, I_t)$ is modified as
\begin{align} 
W_s K(I_s, I_t) Z_t \approx \begin{bmatrix}
0 & 0 \\
0 & K(\tilde{I}_s, \tilde{I}_t)
\end{bmatrix},\label{eq:WKZ_tau}
\end{align}
which can effectively remove the redundant portion of $K(I_s, I_t)$. The effect of this scaling can be seen for the first cluster in Fig. \ref{fig:h2algo}(d). As we skeletonize a cluster $\tau$, we replace its original index set $I_\tau$ with those selected by the ID: $\tilde{I}_\tau$ (see Line \ref{line:skeleton_leaf} of Algorithm \ref{alg:H2-sketch}). This process can be performed in parallel for all clusters at the leaf level, leading to the leaf level being skeletonized as in \ref{fig:h2algo}(e). 

Since $W_\tau$ and $Z_\tau$ are block unit-triangular matrices, inverting them involves simply flipping the sign of the interpolation matrix, giving us the approximation of the block
\begin{align} 
    K(I_s, I_t) & \approx \begin{bmatrix}
    I & T^T_s\\ 0 & I
\end{bmatrix} \begin{bmatrix}
        0 & 0 \\
        0 & K(\tilde{I}_s, \tilde{I}_t)
    \end{bmatrix}
    \begin{bmatrix}
    I & 0 \\ T_t & I
\end{bmatrix} \nonumber\\
 &=
\begin{bmatrix}
    T^T_s\\ I
\end{bmatrix} 
         K(\tilde{I}_s, \tilde{I}_t)
    \begin{bmatrix}
    T_t & I
\end{bmatrix}=U_s K(\tilde{I}_s, \tilde{I}_t) U_t^T\label{eq:WKZ_tau_inv}
\end{align}
For each cluster $s$ and $t\in\mathcal{F}_s$, the coupling matrix is computed by directly evaluating the matrix entries at the skeletonization indices of those clusters: $B_{s,t} = K(\tilde{I}_s, \tilde{I}_t)$. Note that for $t\notin\mathcal{F}_s$, $B_{s,t}$ is not explicitly formed and will be sketched at higher levels.  

\subsubsection{Construction at higher levels}

To continue the skeletonization process beyond the leaf level, we must first ensure that we have samples of the remaining admissible skeletonized portion of the matrix. At the end of the skeletonization process for the leaf level, the matrix $K$ has been transformed by two block diagonal matrices $W^1$ and $Z^1$ where the diagonal blocks are the previously defined $W_\tau$ and $Z_\tau$ matrices, respectively. We define the transformed matrix $K$ at level $l$ to be $K^l = W^l K^{l-1} Z^l$ with $K^0 = K$. To continue the skeletonization at level $l$, we need to extract samples $Y^l$ of $K^l$ relying only on the data from the input and output vectors of the previous levels $Y^{l-1},\Omega^{l-1}$ with $Y^0 = Y = K \Omega$ and $\Omega^0 = \Omega$. First, we transform the random input vectors using $(Z^l)^{-1}$ to obtain the next set of input vectors $\Omega^1 = (Z^l)^{-1}\Omega^{l-1}$. Using these input vectors on $K^l$ gives us:
\begin{align}
    Y^l &= K^l \Omega^l = W^l K^{l-1} Z^l (Z^l)^{-1}\Omega^{l-1} \nonumber\\
    &= W^l K^{l-1} \Omega^{l-1} = W^l Y^{l-1}\label{eq:Yl}
\end{align}
This allows us to represent the samples at level $l$, $Y^l$, as a transformation of the samples of the previous level $Y^{l-1}$. In fact, one can see from (\ref{eq:WKZ_tau}) that $K^l$ has been significantly sparsified compared with $K^{l-1}$ and one doesn't need to form $Y^l$ and $\Omega^{l}$ in full. Instead, substituting (\ref{eq:WKZ_tau_inv}) into (\ref{eq:Yl}) reveals that one only needs to compute subvectors of $Y^l$ and $\Omega^{l}$ once blocks at level $l-1$ has been skeletonized. 

We can see this for the leaf level and higher levels as the follows: (a) At the leaf level, we can compute them as $Y^{2}_\tau=Y_\tau^{\mathrm{loc}}(J_\tau,:)$ (Line \ref{line:update_Y_leaf} of Algorithm \ref{alg:H2-sketch}) and $\Omega^{2}_\tau=U_\tau^T{\Omega}^1_\tau$ (Line \ref{line:update_Omega_leaf} of Algorithm \ref{alg:H2-sketch}). (b) At a higher level $l$, the samples and random vectors for cluster $\tau$ are formed as  ${\Omega}^l_\tau=\begin{bmatrix}
	\Omega^l_{\nu_1}\\
	\Omega^l_{\nu_2}
\end{bmatrix}$ and ${Y}^l_\tau=\begin{bmatrix}
	Y^l_{\nu_1}\\
	Y^l_{\nu_2}
\end{bmatrix}$, respectively. Here $\nu_1$ and $\nu_2$ are the children of $\nu$. The contribution to $Y_\tau^l$ for $\mathcal{F}_{\nu_1}$ and $\mathcal{F}_{\nu_2}$ is subtracted out as $Y_\tau^{\mathrm{loc}}=Y_\tau^l-\begin{bmatrix}
\sum_{b\in\mathcal{F}_{\nu_1}}B_{\nu_1,b}\Omega^l_{b} \\
\sum_{b\in\mathcal{F}_{\nu_2}}B_{\nu_2,b}\Omega^l_{b}
\end{bmatrix}$ (see Line \ref{line:yloc_high} of Algorithm \ref{alg:H2-sketch}). Note that the coupling matrices $B_{\nu_1,b}$ and $B_{\nu_2,b}$ have been explicitly formed at the previous level. As an example, see Fig. \ref{fig:h2algo}(f) where $B_{\nu_1,b}$ and $B_{\nu_2,b}$ have been marked in red, and \ref{fig:h2algo}(g) for the blocks that contribute to $Y_\tau^{\mathrm{loc}}$. Then the transfer matrices in (\ref{eqn:nested_basis}), $E_{\nu_1}$ and $E_{\nu_2}$, as well as the skeleton indices $\tilde{I}_{\tau}$	 can be computed from the row ID of $Y_\tau^{\mathrm{loc}}$ at Line \ref{line:ID_high} of Algorithm \ref{alg:H2-sketch}. Note that a row permutation matrix has been ignored at Line \ref{line:ID_high}. Now we can compute the subvectors of $Y^{l+1}$ and $\Omega^{l+1}$ as $Y^{l+1}_\tau=Y_\tau^{\mathrm{loc}}(J_\tau,:)$ and $\Omega^{l+1}_\tau=\begin{bmatrix}E_{\tau_1}^T & E_{\tau_2}^T\end{bmatrix}{\Omega}^l_\tau$, respectively. See \ref{fig:h2algo}(h) for blocks that correspond to $Y^{l+1}$ and $\Omega^{l+1}$. Finally, for each cluster $b\in\mathcal{F}_\tau$, the coupling matrix is computed by directly evaluating the matrix entries at the skeletonization indices of those clusters: $B_{\tau,b} = K(\tilde{I}_\tau, \tilde{I}_b)$.	  

Algorithm \ref{alg:H2-sketch} represents the extension of the sketching-based construction algorithm for the HSS matrix \cite{martinsson2011fast} to strongly-admissible $\mathcal{H}^2$ matrices. Therefore, we claim that computational complexity and error behavior of Algorithm~\ref{alg:H2-sketch} can be analyzed by extending corresponding analyses in`\cite{martinsson2011fast}. 
We leave the detailed analyses as future work. It is worth noting that HSS typically reveals large, non-constant ranks causing superlinear computational complexity for higher-dimensional problems, but the $\mathcal{H}^2$ matrix allows for linear CPU and memory complexity with small ranks. 
One can readily see that Algorithm \ref{alg:H2-sketch} is an $O(N)$ algorithm assuming the rank $r$ is a small constant (more precisely, $r=O(\log 1/\epsilon)$ and we assume the tolerance is fixed in this paper), as it only requires $O(r^2N)$ time to generate samples $Y$ and direct evaluation of $O(rN)$ matrix entries.

\subsection{Construction With Adaptive Sampling}
\label{sec:adaptive_sample_construction}
The number of samples needed to satisfy a specific relative error threshold $\epsilon$ for the construction of a hierarchical matrix is typically not know beforehand. A few changes to the fixed rank algorithm are needed to support adaptive construction. First, before performing ID to determine the skeletonization indices, we ensure that the current set of samples contain enough data to approximate the node (see the convergence test at Lines \ref{line:convergence_leaf} and \ref{line:convergence_high} of Algorithm \ref{alg:H2-sketch}). This can be achieved by computing the QR decomposition of a node's set of sample vectors $Y_\tau^{\mathrm{loc}}$ and examining the smallest absolute value of the diagonal of the triangular factor. If this value is less than an absolute error threshold $\epsilon_{abs}$, then we consider the node converged. To support a relative threshold, an approximate norm of the matrix can be provided via sketching and the absolute threshold $\epsilon_{abs}$ would simply be the product of the relative threshold $\epsilon$ and the norm. If not converged, we add more samples by $\bar{Y}=K_{blk}(\bar{\Omega})$ with a new random matrix $\bar{\Omega} \in \mathbb{R}^{N\times d}$. These new samples are used to update $\bar{Y}=K_{blk}(\bar{\Omega})$ and $\Omega_\tau^l$ represented by the updateSamples function at Lines \ref{line:sampleupdate_leaf} and \ref{line:sampleupdate_high}.  When all nodes within a level have converged, then we can stop adding samples and move on to ID.

\section{GPU Implementation}
\label{sec:gpu_implementation}
We describe the proposed GPU implementation of the adaptive construction method of Algorithm \ref{alg:H2-sketch} in Section \ref{sec:adaptive_gpu_construction}, followed by a performance analysis in Section \ref{sec:gpuoverall}. 

% In addition, when the operand $K$ consists of an existing $\mathcal{H}^2$ matrix, such as low-rank update to a $\mathcal{H}^2$ matrix, Algorithm \ref{alg:H2-sketch} needs to extract matrix entries directly from the existing $\mathcal{H}^2$ matrix, which is detailed in Section \ref{sec:h2gen}. 

\subsection{Adaptive GPU Sketching}
\label{sec:adaptive_gpu_construction}
In this subsection, we explain the GPU implementation of the proposed algorithm in Algorithm \ref{alg:H2-sketch} in detail. First, we note that launching a kernel on the GPU incurs an overhead that can dominate an application's runtime if the kernel has a small compute workload. Likewise, individual memory allocation for each small kernel can hinder GPU performance. While $\mathcal{H}^2$-matrices provide asymptotically optimal storage and algorithmic complexity, they consist of small dense blocks whose individual kernel launch and memory allocation become impractical. Therefore, a more nuanced approach is required to achieve high performance on GPUs. 

In our proposed implementation, the nodes of the trees are stored contiguously level by level to expose the parallelism in each level of the tree. Then most operations are split into two phases: a marshaling phase where data from the flattened trees relevant to the operation is gathered and a batched execution phase where batch routines carry out all operations using a single kernel call. Unless otherwise stated, the batch count is set to the number of nodes of a given level. The comments in blue-green in Algorithm \ref{alg:H2-sketch} indicate the operations (or loops) that are implemented with GPU kernels. The marshaling routines are executed on the device using the Thrust library~\cite{bell2011thrust} %\ylnote{need a reference for Thrust} 
while the majority of the batched routines are provided by the KBLAS
\cite{abdelfattah2016kblas}
and MAGMA \cite{hdtld15} libraries. Note that most of the batched operations involve non-uniform-sized matrices as the cluster sizes and ranks are not constant. 
One advantage of this approach is that the same code, with trivial modification, can run on either CPUs or GPUs. This is due to that Thrust has multiple parallel CPU backends and the batched routines can be easily implemented on the CPU using parallel OpenMP loops around single threaded BLAS and LAPACK routines. We remark that, when one executes the proposed algorithm on the GPU, all the data, the black-box function $K_{blk}(\cdot)$ and the entry evaluation function fully reside on GPUs. 

%Also, as the workflow of Algorithm \ref{alg:updatesample} resembles that of Algorithm \ref{alg:H2-sketch}, we only explain the GPU implementation of Algorithm \ref{alg:H2-sketch} in what follows. 

We first modify the inputs of Algorithm \ref{alg:H2-sketch} such that instead of a function to evaluate any subblock $K_{s,t}$ on the GPU, we require a function to evaluate a batch of subblocks on the GPU. We call this function batched entry generator and it's invoked to evaluate all $D_{\tau,b}$ or $B_{\tau,b}$ at a given level $l$ with a single kernel launch (see Lines \ref{line:D} and \ref{line:B} marked by \textit{batchedGen}). Next, the random matrices $\Omega$ or $\bar{\Omega}$ are generated in a single kernel and supplied to $K_{blk}(\cdot)$ to produce $Y$ or $\bar{Y}$ on the GPU (see Lines \ref{line:rand}, \ref{line:rand_bar_leaf}, \ref{line:rand_bar_high} marked by \textit{batchedRand}). To avoid large amounts of small memory allocations, the total amount needed per level is first determined using a Thrust parallel prefix sum on the block dimensions and then allocated in a single allocation per operation.\\
\indent When computing $Y_\tau^{\mathrm{loc}}$ from $Y_\tau^l$ to ensure that the samples only include the influence of the admissible blocks as in Fig. \ref{fig:h2algo}(c) and \ref{fig:h2algo}(g), we use a non-uniform batched block sparse row (BSR) matrix multiplication routine. For example, the BSR matrix would include the red dense blocks in Fig. \ref{fig:h2algo}(f). Since no GPU implementation for non-uniform blocks in a BSR matrix product currently exists, we take advantage of the sparsity constant described in Section \ref{sec:preliminaries} to split the operation into at most $C_{sp}$ kernels each performing a batched non-uniform matrix-matrix multiplication using MAGMA. Each kernel works on marshaled data in a way that allows us to update the output vectors in parallel without resorting to atomic operations; that is to say that only one block from each row will be involved in each kernel launch, and since we have at most $C_{sp}$ such kernels, the kernel launch overhead should not impact performance. \\
\indent After obtaining a set of samples of the admissible part of the matrix, the convergence test checks if all nodes have converged to satisfy the error threshold $\epsilon_l$ using the method described in Section \ref{sec:adaptive_sample_construction}. If the current set of samples prove to be insufficient, additional samples and input vectors are generated. The updateSamples function at Lines \ref{line:sampleupdate_leaf} and \ref{line:sampleupdate_high} sweeps any new samples and input vectors up the tree until it reaches the current level. \\
\indent Once we've acquired a sufficient number of samples $Y_\tau^{\mathrm{loc}}$, a batched row ID determines the skeletonization indices. As the row ID is implemented via the column ID on the matrix transpose. The list of samples are first accumulated using a batch transpose to allow for more efficient memory access patterns on the GPU for a column pivoted QR. See Lines \ref{line:ID_leaf} and \ref{line:ID_high} marked by \textit{batchedID}. \\
%The produced indices are sorted for convenience of the batched entry generator (as we show in Section \ref{sec:h2gen}) and represent a relatively insignificant computational load since each index set is small. See Lines \ref{line:ID_leaf} and \ref{line:ID_high} marked by \textit{batchedID}. \\
\indent Finally, the input vectors $\Omega_\tau^l$ are upswept to the next level using a batched matrix multiplication (see Lines \ref{line:update_Omega_leaf} and \ref{line:update_Omega_high} marked by \textit{batchedGemm}) and the samples $Y_\tau^{\mathrm{loc}}$ are upswept by first swapping the columns of the previously transposed samples which are then transposed again to their row skeletonized form $Y_\tau^{l+1}$ (see Lines \ref{line:update_Y_leaf} and \ref{line:update_Y_high} marked by \textit{batchedShrink}).  

\begin{figure*}[th!]
	\centering
	\includegraphics[width=\textwidth]{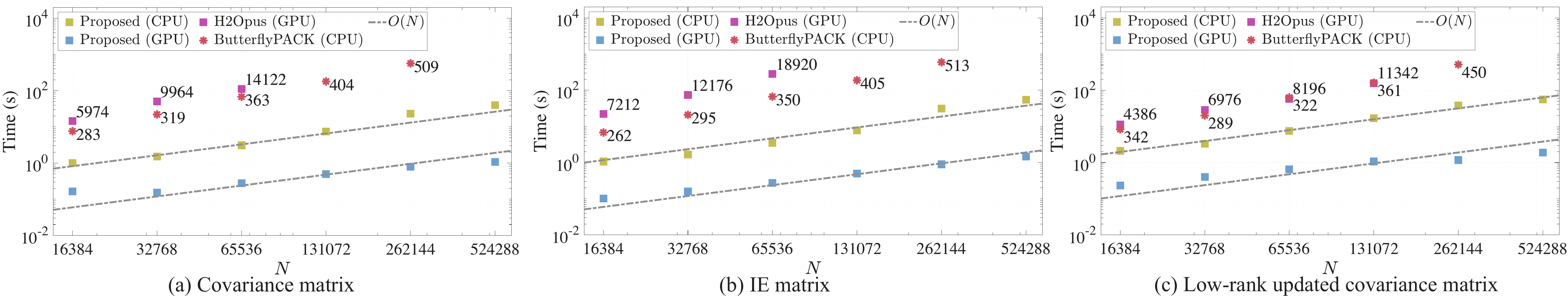}
	\caption{The time of the CPU and GPU implementations of Algorithm \ref{alg:H2-sketch} for the covariance and IE matrices as well as the covariance matrix updated with a rank $32$ low-rank product. Also shown on the time plots is the top-down construction using H2Opus and ButterflyPACK with its data points labeled with the total samples taken.}\label{fig:cpu}
\end{figure*}

\begin{figure*}[th!]
	\centering
	\includegraphics[width=0.9\textwidth]{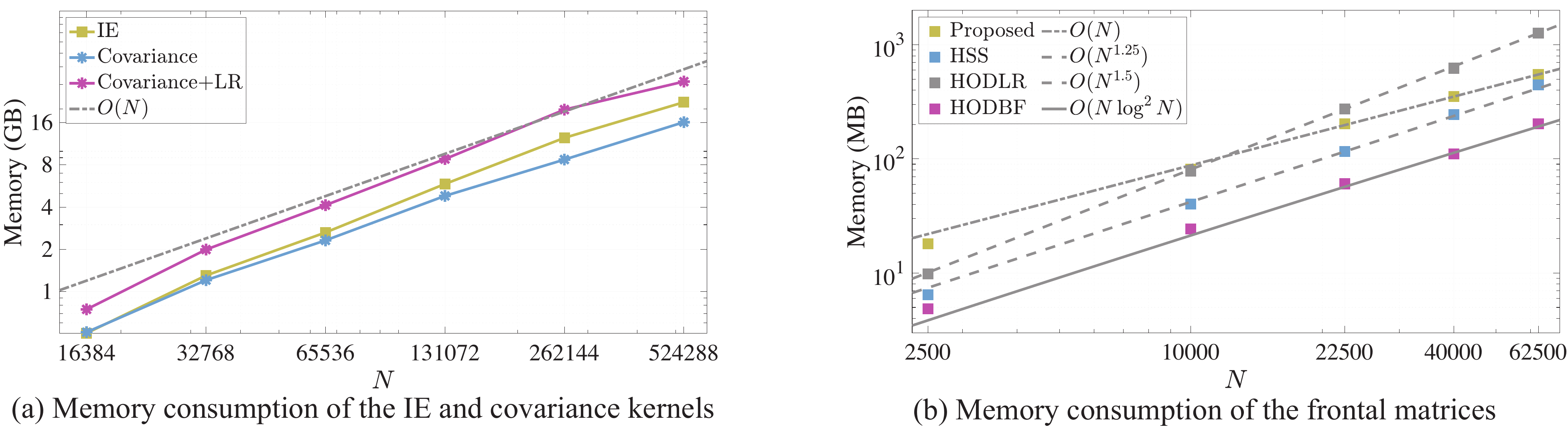}
	\caption{(a) The memory of Algorithm \ref{alg:H2-sketch} for the covariance and IE matrices. (b) The memory of Algorithm \ref{alg:H2-sketch} and a few other sketching-based algorithms in STRUMPACK for the frontal matrices. }\label{fig:mem}
\end{figure*}

\subsection{Performance Analysis }
\label{sec:gpuoverall}
Here we provide a brief analysis of the parallelization performance of Algorithm \ref{alg:H2-sketch} on the GPU. 
As described in Section \ref{sec:gpu_implementation}, the operations implemented on GPU are batchedRand, batchedBSRGemm, batchedID, and batchedShrink, and the inputs (executed on the GPU as well) are the black-box function $K_{blk}(\cdot)$ and batchedGen. Although Section \ref{sec:gpu_implementation} and the numerical results only involve single-GPU implementation, here we add a few notes regarding the potential extension of our algorithm to multiple GPUs.  

All the batched operations have a batch count set to the number of nodes at that level. The batch count decreases from $O(N)$ at the leaf level to (at most) $O(1)$ at the highest level. When the ranks have the same order of magnitudes across the levels, the computation workload per node $\tau$ remains relatively constant and hence higher parallel efficiency can be achieved for lower levels. Note that $\mathcal{H}^2$ is a $O(N)$-complexity algorithm dominated by the lower level operations, good overall parallel efficiency can be achieved for these operations. It is also worth mentioning that our batched algorithm requires only $L=O(\log N)$ kernel launches, which is a very small cost compared with the total $O(N)$ computational cost. In fact, the overhead in kernel launches is negligible compared with the total execution time, particularly for large $N$.   

For multiple GPUs, the batch count becomes roughly the number of nodes per level divided by the number of GPUs. Moreover, all the aforementioned batched operations do not require inter-GPU communication except for batchedBSRGemm, which requires communication of the input vectors $\Omega$. Also Line \ref{line:merge} of Algorithm \ref{alg:H2-sketch} may require gathering the vectors from two GPUs into one.         

%In principle, our proposed algorithm is oblivious to the implementation details of the input functions, i.e., $K_{blk}(\cdot)$ and batchedGen, as they are provided by the users. However, when the application is to construct a updated $\mathcal{H}^2$-matrix from an existing $\mathcal{H}^2$ matrix, one may need to provide a function $K_{blk}(\cdot)$ to perform matrix-matrix multiplication with $\mathcal{H}^2$-matrices on GPUs (e.g., via multiplication algorithms in \cite{zampini2022h2opus}), and a function batchedGen to extract entries from $\mathcal{H}^2$-matrices on GPUs (see Section \ref{sec:h2gen}). Although achieving high GPU efficiency for the batched entry extraction is challenging, we note that batchedGen doesn't represent the computation bottleneck of the overall algorithm (see Section \ref{sec:profile})

\section{Numerical Evaluation}\label{sec:numerical_result}

\label{sec:application_and_performance}
In this section, we analyze the performance of Algorithm \ref{alg:H2-sketch} on three different problems on CPUs and GPUs. We compare the memory, runtime and accuracy of our algorithm with other existing high-performance implementations of sketching-based strongly-admissible hierarchical matrix construction algorithms, including the GPU implementation of the top-down $\mathcal{H}^2$ algorithm \cite{lin2011fast} from the H2Opus library~\cite{zampini2022h2opus} and the distributed-memory CPU implementation of the top-down $\mathcal{H}$ algorithm \cite{levitt2022randomized} from the ButterflyPACK (v3.2.0) library~\cite{liu2018butterflypack}. To the best of our knowledge, these are the only publicly available packages supporting sketching-based construction of strongly-admissible hierarchical matrices (i.e., $\mathcal{H}^2$  or $\mathcal{H}$). Our proposed GPU implementation of Algorithm \ref{alg:H2-sketch} and the reference algorithm in H2Opus are executed on an 80GB A100 GPU available on Perlmutter GPU nodes. Our proposed CPU implementation Algorithm \ref{alg:H2-sketch} uses OpenBLAS\footnote{https://github.com/OpenMathLib/OpenBLAS} routines within OpenMP parallel loops for the batched operations and Thrust with the OpenMP backend for the data marshaling, which is executed using 64 OpenMP threads of an AMD EPYC 7763 processor available on Perlmutter GPU nodes. The reference algorithm in ButterflyPACK is executed on the same AMD processor using 64 MPI ranks. In addition to ButterflyPACK and H2Opus, we also consider comparison with sketching-based weakly-admissible hierarchical matrix construction algorithms implemented in STRUMPACK (v8.0.0) \cite{osti_1328126}. 

\subsection{Test Problems}
Throughout the paper, we consider three applications of the proposed $\mathcal{H}^2$ construction algorithm. For the first application, we look at the construction of spatial statistics covariance matrices for a 3D Gaussian spatial process on a uniform 3D distribution of points in a cube and use an exponential kernel with correlation length $l = 0.2$:
\begin{equation}
    K(x, y) = e^{-\frac{|x - y|}{l}}\label{eq:cov}
\end{equation}
For the second application, we consider the construction of the discretized volume integral equation (IE) operator for the Helmholtz equation among a uniform 3D distribution of points in a cube and the IE operator is  
\begin{equation}
    K(x, y) = \frac{\cos{(k|x - y|)}}{|x - y|},~ x\neq y\label{eq:ie}
\end{equation}
with $k$ fixed to be 3. 
For these two applications, we use the fast $\mathcal{H}^2$-matrix-vector product from the H2Opus library~\cite{zampini2022h2opus}
as the black box input function $K_{blk}(\cdot)$ and the direct implementation of (\ref{eq:cov}) and (\ref{eq:ie}) in batchedGen. For the reference CPU implementation from ButterflyPACK, we use the $\mathcal{H}$ representation for $K_{blk}(\cdot)$ and implement batchedGen on CPUs.   

For the third application, we extract frontal matrices of varying sizes in full from the multifrontal factorization of a uniform-grid discretized 3D Poisson problem. We compare the performance of the proposed algorithm with other sketching-based algorithms implemented in STRUMPACK \cite{osti_1328126} (e.g. HSS \cite{martinsson2011fast}, HODLR \cite{lin2011fast} and HODBF).  

In addition to construction of the $\mathcal{H}^2$ matrix from these kernels, we also consider the updating an existing $\mathcal{H}^2$ representation of the covariance matrix with an additional low-rank product using the proposed algorithm. This is commonly encountered during the LU decomposition of hierarchical matrices or in the multifrontal factorization of sparse matrices. We use the fast $\mathcal{H}^2$-matrix-vector product from H2Opus (and fast low-rank multiplication) to perform $K_{blk}(\cdot)$, and an algorithm that extracts entries from the given $\mathcal{H}^2$ and low-rank representations to perform batchedGen. 

The cluster tree is constructed as a KD-tree with a leaf size of 64-256 and a dual tree traversal of the cluster tree constructs the matrix tree. We measure the approximation relative error $\frac{|K_{\mathrm{comp}} - K|}{|K|}$ using a few iterations of the power method to approximate the 2-norm of the difference between the constructed hierarchical matrix and the provided sampler $K_{blk}(\cdot)$.

\begin{figure*}[th!]
	\centering
	\includegraphics[width=\textwidth]{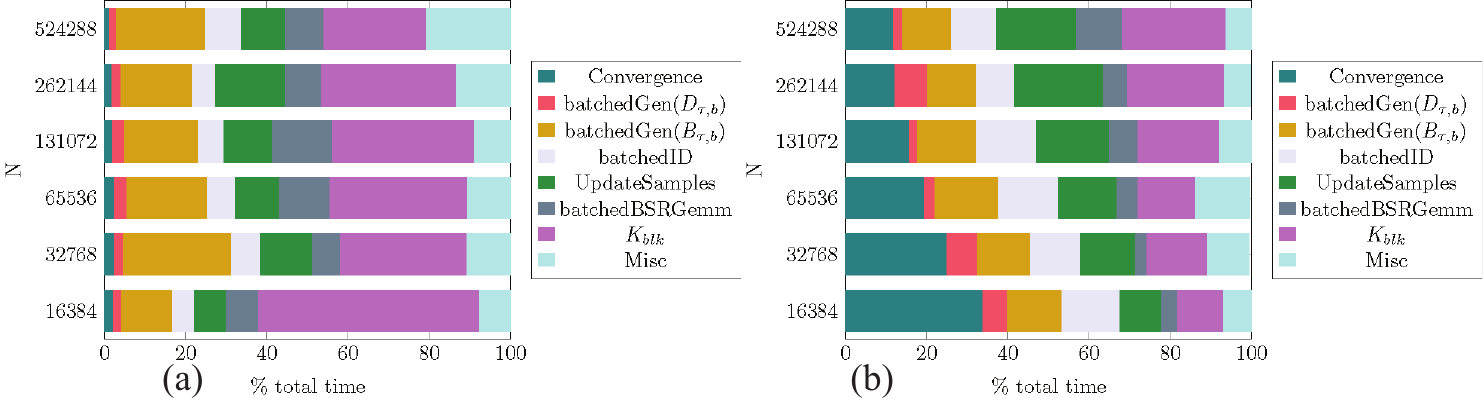}
	\caption{A breakdown of the construction time by percentage of time taken by each phase on (a) CPU and (b) GPU for varying problem sizes of the 3D covariance matrix.}
	\label{fig:cpu_gpu_profile}
\end{figure*}

\begin{table*}[th!]
	\begin{centering}
		\begin{tabular}{|l|l|c|c|c|c|c|c|c|}
			\hline
			& & Time  & Rank range & Memory   & Total samples  	& Sample block size & Leaf Size & Relative Error        \\ \hline
			Covariance  &fixed sample & 0.860	& 29-55		 & 22.939	& 256		& 256		 & 256		 & 4.139360e-08	\\ \cline{2-9} 
			&adaptive       & 0.302	& 45-55		 & 22.956	& 64		& 32		 & 256		 & 5.374687e-07 \\ \cline{2-9}
			& fixed sample & 0.348	& 32-55		 & 13.906	& 128		& 128		 & 128		 & 8.404928e-08 \\ \cline{2-9}
			&adaptive	& 0.246	& 42-55		 & 13.930	& 64		& 32		 & 128		 & 5.968764e-07 \\ \hline
			IE     &fixed sample   & 0.897	& 33-66		 & 23.437	& 256		& 256		 & 256		 & 2.603808e-08 \\ \cline{2-9}
			&adaptive & 0.428	& 36-66		 & 23.446	& 96		& 32		 & 256		 & 1.120914e-07 \\ \cline{2-9}
			&fixed sample	& 0.517	& 35-66		 & 14.998	& 128		& 128		 & 128		 & 6.461307e-08 \\ \cline{2-9}
			&adaptive & 0.392	& 51-67		 & 15.081	& 96		& 32		 & 128		 & 1.614852e-07 \\ \hline
		\end{tabular}
		\caption{The effect of varying the leaf size and sample block size on the memory consumption and ranks of the constructed matrix as well as runtime and approximation error with a threshold of $10^{-6}$ for the 3D problems of size $N=2^{18}$.} %\ylnote{Wajih, please add a sentence about the accuracy of H2OPUS}} 
		\label{tab:adaptive_sampling}
	\end{centering}
\end{table*}

\subsection{Computational Complexity}
For the IE and covariance kernels, we examine the overall performance of the algorithm to verify the optimal complexity of the construction in time and memory. Fig. \ref{fig:cpu}(a) and \ref{fig:cpu}(b) show the construction time including sampling and entry generation for various covariance and IE matrices respectively on the CPU and GPU while figure \ref{fig:cpu}(c) shows the same statistics for compressing the sum of $\mathcal{H}^2$ representation of the covariance matrix and a rank-$32$ low-rank product into a new $\mathcal{H}^2$ matrix. The memory usage of the proposed algorithm is shown in Fig. \ref{fig:mem}(a). We also show the GPU top-down construction time from the H2Opus library.
The target matrices were constructed to an error threshold of $10^{-6}$ with an admissibility parameter $\eta = 0.7$, $256$ initial samples and a leaf size of $64$, while the input H2Opus matrices was constructed to a looser threshold of $10^{-5}$ as the implementation within H2Opus uses Cholesky QR for the orthogonalization of samples and thus it is difficult to reliably construct matrices with tight thresholds. The input ButterflyPACK matrices was also constructed with a threshold of $10^{-5}$. 

Our construction algorithm clearly exhibits the expected optimal runtime complexity with our GPU implementation showing speedups of up to $13\times$ over our CPU implementation, up to $660\times$ over ButterflyPACK's CPU implementation, and over $1000\times$ faster than H2Opus' GPU implementation. (see Fig. \ref{fig:cpu}(a)-(c)). Note that H2Opus runs out of memory on the problems larger than $65536$. Fig. \ref{fig:mem}(a) shows the expected linear growth of the memory consumption of the constructed matrices for the three test problems. 

It's worth mentioning that our algorithm requires $O(1)$ (more precisely 256 for all data points in Fig. \ref{fig:cpu}) number of random vectors. In stark contrast, the algorithm in ButterflyPACK \cite{levitt2022randomized} requires $O(\log N)$ random vectors (ranging from 262 to 513 in Fig. \ref{fig:cpu}(a)-(c)). Moreover, H2Opus's implementation requires a temporary weak-admissible representation (HODLR), hence requires much more number of random vectors (up to 18920) for 3D problems, causing the code to memory crash for larger problems. In short, we remark that the proposed algorithm requires significantly less random samples particularly when the problem size $N$ increases. This dramatic reduction in the number of random vectors, i.e., the time spent in $K_{blk}(\cdot)$, contributes most to the aforementioned speedups comparing with ButterflyPACK or H2Opus.  

Also, note that the largest problem size $N=524288$ of our GPU implementation is limited by the fact that we need to store both $K_{blk}(\cdot)$ (consisting of an existing $\mathcal{H}^2$ matrix) and the constructed $\mathcal{H}^2$ matrix on the single GPU. Given that one A100 GPU with 80GB memory is used, the $\mathcal{H}^2$ matrix can consume at most approximately 40GB memory. To handle larger problem sizes, a multi-GPU implementation will be considered as our future work.  

For the frontal matrices, we only show the memory usage of different algorithms as the sketching operator is a full $N\times N$ matrix. This also limits the largest matrix sizes we can test. We leave the full integration of the proposed algorithm into multi-frontal sparse direct solvers as a future work. Fig. \ref{fig:mem}(b) shows the memory usage of the proposed algorithm, HSS \cite{martinsson2011fast}, HODLR and HODBF. Clearly, our algorithm achieves the optimal $O(N)$ memory usage. However, note that the other three algorithms are weak-admissibility-based and their prefactors can be much smaller.

\subsection{Profiling Results}\label{sec:profile}
Fig. \ref{fig:cpu_gpu_profile} breaks down the runtime into the major components of the construction algorithm on the CPU and the GPU. The convergence test, where the batched QR decomposition comprises the majority of the work, represents a significantly smaller portion of the overall runtime on the CPU compared to the GPU. This is primarily caused by the batched QR implementation within KBLAS favoring larger batch sizes and smaller matrices with it's unblocked algorithm that only assigns threads to work on one column at a time. This is especially apparent for the smaller problem sizes where there isn't enough work to saturate the GPU. As the problem size increases, the overall portion of time spent in the convergence test on the GPU starts to shrink.
The dense and coupling $\mathcal{H}^2$ entry generation seems to perform well on both the CPU and GPU, taking between 10-15\% of the total runtime on the CPU and 15-20\% on the GPU. Since we only perform the pivoted QR decompositions after we've determined that the number of samples are sufficient, the ID phases only take between 5-10\% of the runtime. 
On both CPU and GPU, the majority of time is spent in the BSR matrix multiplication and the sampling phases, both of which are heavily matrix-matrix multiplication dependent, an operation that is particularly well suited for GPU execution. The miscellaneous section includes mostly workspace allocations which can be optimized in the future.

\subsection{Efficacy of Adaptive Sampling}
Finally, we demonstrate the effects of the adaptive sampling on a fixed 3D problem of size $N=2^{18}$ by varying both the leaf size and the sampling block size $d$. Table \ref{tab:adaptive_sampling} shows the GPU results for both the covariance and IE matrices for leaf sizes of $128$ and $256$ and sampling block sizes equal to the leaf size and fixed at $32$. The lower leaf sizes lead to lower overall memory consumption and lower construction times while the fixed sampling block sizes lead to overall lower execution times albeit with a lower resulting accuracy and slightly higher ranks on the higher levels. This is likely due to the simple error compensation scheme not fully accounting for the approximation errors as we sweep up the tree, though the measured error is still within the threshold of $10^{-6}$.

\section{Conclusion}
\label{sec:conclusion}
This paper presents a GPU algorithm and implementation of a novel linear-complexity bottom-up sketching-based algorithm for constructing a $\mathcal{H}^2$ matrix. 
The proposed construction algorithm requires both a black-box sketching operator and an entry evaluation function, both of which are accelerated by batched GPU implementations. When applied to covariance matrices, volume IE matrices and $\mathcal{H}^2$ update operations, our proposed GPU implementation achieves up to $13\times$ speedup over our CPU implementation, and up to $1000\times$ speedup over an existing GPU implementation of the top-down sketching-based algorithm from the H2Opus library. Moreover, the proposed algorithm is capable of handling covariance/IE matrices with sizes up to $N=524288$ using less than $30$ GB GPU memory and we expect the algorithm can go up to $N=1.5$ million on a single 80GB A100 GPU with further code optimizations in the future. 

In addition to pushing the limit of the proposed algorithm, we also plan to investigate the GPU implementation of the inversion of the $\mathcal{H}^2$ matrix \cite{Ma2019} and the GPU implementation of other fully sketching-based construction algorithms such as \cite{Yesypenko,levitt2022randomized}, as well as the full integration of the proposed algorithm into sparse multifrontal solvers. 

%\ylnote{Not sure if it's worth mentioning here, but there are many other things to be done including support for (1) complex matrices, (2) support for non-symmetric matrices, (3) support for multi-GPU, (4) factorization, (5) CB compression without entry evaluation etc. }

% \todo[inline]{Complete}

% GPU implementations of $\mathcal{H}^2$ construction and inversion are linear complexity. Maybe single-GPU implementation are good enough in practice? No need for massive parallelism? Add some numbers from table II. 

% Add punchline here as well. 

% Our future work includes the following.
% \begin{itemize}
% \item We plan to investigate the GPU implementation of the fully black-box $\mathcal{H}^2$ construction algorithm % \cite{Yesypenko} and compare with our partially black-box-based algorithm. 
% \item We plan to look into the GPU implementation of the factorization of the $\mathcal{H}^2$ matrix algorithm as well.
% \end{itemize}

\section*{Acknowledgment}
This research was supported by the Exascale Computing Project (17-SC-20-SC), a collaborative effort of the U.S. Department of Energy Office of Science and the National Nuclear Security Administration.

% \begin{verbatim}
%   \begin{acks}
%   ...
%   \end{acks}
% \end{verbatim}

% \section{Appendices}
% \begin{verbatim}
%   \appendix
% \end{verbatim}


\bibliographystyle{IEEEtran}
\bibliography{sample-base}

\end{document}